\pdfoutput=1
\documentclass[journal]{IEEEtran}
\ifCLASSINFOpdf
  \usepackage[pdftex]{graphicx}
\else
\fi
\usepackage{cite}
\usepackage{amsmath}
\usepackage{amsfonts} 
\newcommand{\sgn}{\operatorname{sign}} 
\usepackage{algorithmic}
\usepackage{array}
\usepackage{color}
\usepackage{subfigure}
\usepackage[justification=centering]{caption}
\usepackage{mdwtab}
\begin{document}
\title{Robust Time-Frequency Reconstruction\\ by Learning Structured Sparsity}

\author{Lei~Jiang, Haijian~Zhang, and Lei~Yu 
\\Signal Processing Lab., School of Electronic Information, Wuhan University, China
       \vspace{-.8\baselineskip}
}

\markboth{Time-frequency analysis and its applications}{}
\maketitle

\begin{abstract}
Time-frequency distributions (TFDs) play a vital role in providing descriptive analysis of non-stationary signals involved in realistic scenarios. It is well known that low time-frequency (TF) resolution and the emergency of cross-terms (CTs) are two main issues, which make it difficult to analyze and interpret  practical signals using TFDs. In order to address these issues, we propose the U-Net aided iterative shrinkage-thresholding algorithm (U-ISTA) for reconstructing a near-ideal TFD by exploiting  structured sparsity in signal TF domain. Specifically, the signal ambiguity function is firstly compressed, followed by unfolding the ISTA as a recurrent neural network. To consider continuously distributed characteristics of signals, a structured sparsity constraint is incorporated into the unfolded ISTA by regarding the U-Net as an adaptive threshold block, in which structure-aware thresholds are learned from enormous training data to exploit the underlying dependencies among neighboring TF coefficients. The proposed U-ISTA model is trained by both non-overlapped and overlapped synthetic signals including closely and far located non-stationary components. Experimental results demonstrate that the robust U-ISTA  achieves superior performance compared with state-of-the-art algorithms, and gains a high TF resolution with CTs greatly eliminated even in low signal-to-noise ratio (SNR) environments.
\end{abstract}
\begin{IEEEkeywords}
 time-frequency distribution, compressive sensing, iterative shrinkage-thresholding, deep learning
\end{IEEEkeywords}
\IEEEpeerreviewmaketitle

\section{Introduction}
\IEEEPARstart{D}{uring} the past decades, various time-frequency distributions (TFDs), arising in the field of time-frequency (TF) signal analysis and processing \cite{2013LjubisaStankovic,2016ivBoualem,2018CambridgePatrick,SEJDIC2009153}, have been reported and developed to deal with a wide range of non-stationary signals whose magnitudes and frequencies change over time, such as radar signals \cite{8353500,670330,7073495Zhang}, communication signals \cite{keller2000adaptive,8986251,8026105Amin}, audio signals \cite{luo2019conv,5109766,8894138Hua}, and biomedical signals \cite{tsinalis2016automatic,4801967}. The vast majority of practical signals are non-stationary, thus seeking a near-ideal TF representation of non-stationary signals is of great significance to signal processing applications including filtering \cite{1275667Boashash,7342935Zhang}, localization \cite{1608548Yimin,ZHANG201625,7955025Ouelha}, separation \cite{6553137Stankovic,ZHANG2015141,ZHANG20171}, classification \cite{905863Gillespie,6931399Zhang,BOASHASH2018120}, etc. In this paper, we emphasize on the robust reconstruction of TF domain mixture signal under different signal-to-noise ratio (SNR) conditions.

All the while, short-time Fourier transform (STFT) is the most commonly used linear TFD whose resolution is restricted to fixed sliding window \cite{ZHANG201625}, and the wavelet transform (WT) \cite{daubechies1990wavelet} with adaptive window size  was proposed to attain changeable TF resolution. Though linear TFDs have desirable property, e.g., low implementation complexity, nonlinear TFDs can achieve more concentrated TF energy distributions \cite{127284Hlawatsch}. Typical nonlinear TFDs include polynomial TFDs \cite{4034118Bi,boashash1998polynomial},  Wigner-Ville distribution (WVD) \cite{1018780Hussain,fonoliosa1993wigner} and ambiguity function (AF) \cite{6931399Zhang}, however, they suffer from inner and outer cross-terms (CTs) caused by the nonlinear structure of TFDs, which might plague signal TF analysis and thus need to be reduced. From a post-processing point of view, the reassignment method assigns the average energy in certain domain to the gravity center of energy distributions, e.g., reassigned spectrogram (RSP) algorithm \cite{382394}. Nevertheless, it is still difficult to avoid CTs and TFD distortion especially in low SNR environments. The above difficulty highlights the requirement for improving TF reconstruction performance by breaking the tradeoff between high TF resolution and significant CTs \cite{7401107Zuo,7736146Zuo}.

An increasing number of kernel function based TF analysis methods emerge to seek a compromise between suppressing CTs and preserving high TF resolution \cite{950779,boashash2013time,baraniuk1993signal,boashash2017improved,8868836,saulig2019extraction}. Since auto-terms (ATs) are distributed around the AF domain origin while most of CTs locate far away from the origin, classical Cohen's class distributions \cite{30749Cohen} focus on designing various kernel functions in AF domain to remove CTs. On the one hand, some kernel functions could be signal independent, e.g., B-distribution (BD) \cite{950779} and extended modified B-distribution (EMBD) \cite{boashash2013time}, which utilizes separable kernels to control the smoothing requirements between time and frequency axes. On the other hand,  some kernel functions for special signals may be signal dependent as well, e.g., radially Gaussian kernel (RGK) \cite{baraniuk1993signal} and multi-directional distribution (MDD) \cite{boashash2017improved}, which is specifically designed for  piece-wise linear frequency-modulated signals. Besides, extracting useful information directly from typical TFDs such as WVD would be an alternative strategy to remove CTs \cite{8868836,saulig2019extraction}.

Classical time-frequency dictionaries have been combined with compressive sensing (CS) algorithms to establish sparse representations of non-stationary signals \cite{6361356Gholami}, e.g., matching pursuit (MP) \cite{mallat1993matching} and orthogonal matching pursuit (OMP) \cite{342465}. The continuous development of CS technology brings new solutions to signal analysis and reconstruction by exploiting the sparsity of signal in transform domains \cite{s0003401809092LJ,SEJDIC201822,YU2012259,WU2020107560,stankovic2020ransacbased,flandrin2010time,volaric2017data,7330283,6671387Whitelonis,moghadasian2019sparsely,8868576,bioucas2007new,amin2019sparsity}. 
Flandrin \textit{et al.} \cite{flandrin2010time} proposed the $\ell_{1}$-app algorithm relying on convex optimization to recover  sparse TFDs of signals. The sparse representation of $\ell_{1}$-app is constructed with the 2-dimensional (2D) Fourier transform pair of  quadratic AF and WVD. A small number of samples around the AF domain origin are treated as observation vector since they contain most energy of auto-terms, while their corresponding Fourier coefficients are used to built the measurement matrix. With approximate $\ell_{1}$ sparsity constraint, sparse TFDs with negligible CTs have been reconstructed \cite{flandrin2010time}. However, when the signal's TFD involves complex structure, e.g., crossing and closely-located components, the performance of $\ell_{1}$-app might degrade. Moreover, fixed sampling geometry neglects  the TF correlation structure of the analyzed signal and inner CTs resulting from the nonlinearity. Therefore, adaptive sampling geometry methods have been proposed to improve TF reconstruction performance \cite{volaric2017data,7330283,6671387Whitelonis}. Inspired by MDD \cite{boashash2017improved}, Moghadasian \textit{et al.}  \cite{moghadasian2019sparsely} proposed a signal-dependent sampling geometry algorithm which gains high TF resolution for signals composed of closely-located components. In addition to convex optimization algorithms, iterative algorithms have also been considered due to their low computational complexity and fast convergent rates. Volaric \textit{et al.} \cite{8868576} combined the  fast intersection of the confidence intervals (FICI) rule with a two step iterative shrinkage thresholding algorithm (TwIST) \cite{bioucas2007new}, and proposed the FICI-TwIST algorithm. Since missing data samples are commonly encountered resulting from sensor failure and jammed measurements, Amin \textit{et al.} \cite{amin2019sparsity} considered a real-life scenario and proposed the missing data iterative sparse reconstruction (MI-SR) algorithm based on instantaneous auto-correlation function (IAF) by taking account of practical burst missing data.

Recently, data-driven learning algorithms have been booming with the development of deep learning technology \cite{gregor2010learning,chen2018theoretical,liu2018alista,Alaskar2018,8100870Sun}. Benefiting from a large number of data, most of learning based algorithms pay  attention to learning self-dependency between thresholds and signal amplitudes   so that a one-to-one correspondence between each threshold and signal amplitude is obtained. It is well known that the structured sparsity model going beyond simple sparsity is feasible in many practical scenarios  by exploiting the inter-dependency structure among signal coefficients \cite{eldar2009robust,baraniuk2010model,eldar2010block}.
Wu \textit{et al.} \cite{6900783} proposed a continuous structure Bayesian method to recover signals with missing data. Intuitively, considering the priori knowledge of structured sparsity, e.g., the continuity of signal TF signatures, will give rise to more robust recovery performance.
Consequently, sophisticated TF reconstruction algorithms by learning structured sparsity prior of signal's TFDs are worthy for further investigation.

In this paper, we propose a new deep learning based structured sparse TF reconstruction algorithm, which is hereafter referred to as the U-Net aided iterative shrinkage-thresholding algorithm (U-ISTA). Unlike other iterative algorithms with manual or self-dependent thresholds \cite{7481717,fosson2018biconvex,gregor2010learning,chen2018theoretical,liu2018alista}, the structure-aware thresholds in U-ISTA are learned via the  U-Net classifier \cite{ronneberger2015u} to reveal underlying relationship among the signal sparsity coefficients in TF domain. The U-Net is recognized as an adaptive block involved into the unfolded ISTA, i.e., a one-to-many correspondence between one threshold and adjacent TF coefficients is built up so that the thresholds depend on both amplitude and structure information of signals. Experimental results demonstrate that the proposed end-to-end U-ISTA network greatly eliminates the CTs while obtaining a near-ideal TF representation even for spectrally-overlapped and closely-located signal components in low SNR environments.

The rest of this paper is organized as follows. Section \uppercase\expandafter{\romannumeral2} introduces the proposed U-ISTA. Experiments on various kinds of synthetic signals are presented in Section \uppercase\expandafter{\romannumeral3}. Finally, Section \uppercase\expandafter{\romannumeral4} concludes this paper.

\begin{figure*}[ht]
\centering
\includegraphics[width=7.1 in]{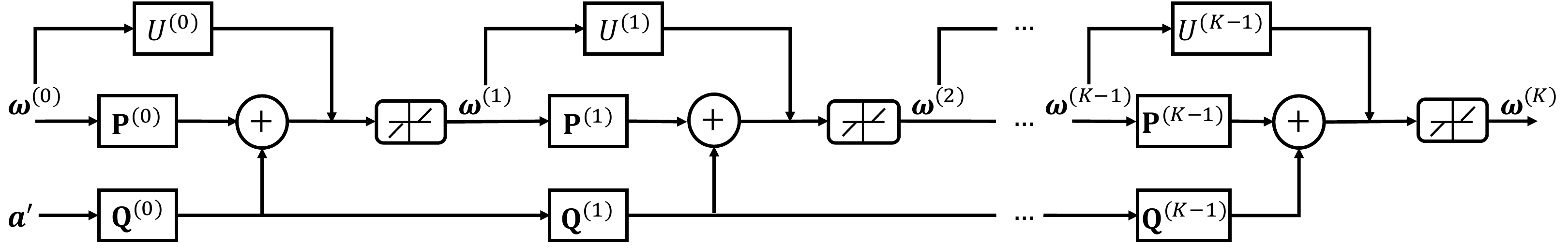}
\vspace{.1\baselineskip}
\caption{The network architecture of the proposed U-ISTA.\label{fig1}}
\end{figure*}

\section{The Proposed U-ISTA} 

\subsection{Problem Statement in TF Domain}
The quadratic WVD of a non-stationary signal $x(t)$ is defined as the Wigner distribution of its analytic associate
\begin{align}
\text{WVD}_{z}(t, f)=\underset{\tau \rightarrow f}{\mathcal{F}}\left\{z\left(t+\frac{\tau}{2}\right) z^{*}\left(t-\frac{\tau}{2}\right)\right\}, \label{eq1}
\end{align}
where $z(t)$ is the analytic associate of $x(t)$. For a signal $x(t)$ containing a single linear frequency-modulated component with infinite duration, its WVD will exhibit ideal signal TF signature and give an exact estimation of signal instantaneous frequency (IF). Nevertheless, practical signals have finite duration, which inevitably results in energy leakage in signal TF domain. In contrast, when the signal $x(t)$ contains multiple  components, the inner and outer CTs  emerge especially when nonlinear components are involved. 

The sparse representation in the proposed U-ISTA is built based on the 2D Fourier relationship between WVD and AF \cite{flandrin2010time}. The signal's AF is interpreted as a joint TF correlation function, and can be calculated by 2D Fourier transform of the signal's WVD 
\begin{align}
\text{AF}_{z}(\nu, \tau)&=\underset{t, f \rightarrow \nu, \tau}{\mathcal{F}_{\text{2D}}}\Big\{\text{WVD}_{z}(t, f)\Big\}.\label{eq2}
\end{align}
The discrete version of analytic signal $z(t)$ is defined as $z(n)$, $n=1,2,...,N$, where $N$ denotes the number of time samples. For brevity, we assume the number of frequency samples is equivalent to that of time samples. Thus, the discrete matrix form in (\ref{eq2}) can be written as
\begin{align}
\mathbf{A} &=\mathbf{D}\mathbf{W}\mathbf{D}^{T},\label{eq3} 
\end{align} 
where $\mathbf{A} \in \mathbb{R}^{N \times N}$, $\mathbf{W} \in \mathbb{R}^{N \times N}$, $\mathbf{D} \in \mathbb{R}^{N\times N}$ is the discrete Fourier transform matrix, and  $T$ denotes the transpose operator. Next, the Kronecker product is used to obtain a convenient representation for the matrix form in (\ref{eq3})
\begin{align}
\boldsymbol{a}&=\mathbf{\Psi}\boldsymbol{\omega},\label{eq4}
\end{align}
where $\boldsymbol{a} \in \mathbb{R}^{N^{2} \times 1}$ denotes the vectorization of the matrix $\mathbf{A}$ formed by stacking the columns of $\mathbf{A}$ into a single column vector,   $\mathbf{\Psi}=\mathbf{D} \otimes \mathbf{D} \in \mathbb{R}^{N^{2}\times N^{2}}$, where $\otimes$ represents the Kronecker product. Similarly,  $\boldsymbol{\omega} \in \mathbb{R}^{N^{2} \times 1}$ denotes the vectorization of the matrix $\mathbf{W}$ formed by stacking the columns of $\mathbf{W}$ into a single column vector.

Since the ATs corresponding to the clean TF signatures mainly distribute around the origin in AF domain, whereas the CTs locate far away from the AF domain origin. This offers an opportunity to remove CTs by appropriately choosing samples in AF domain, which is realized through multiplying a low pass filter $\mathbf{\Phi}(\nu, \tau)$ in AF domain
 \begin{align}
 \text{AF}_{z}^{\prime}(\nu, \tau)&=\text{AF}_{z}(\nu, \tau) \odot \mathbf{\Phi}(\nu, \tau),\label{eq5}
 \end{align}
where $\odot$ denotes element-wise multiplication, thus a new observation vector $\boldsymbol{a}^{\prime}$ can be obtained, and $\mathbf{\Phi}(\nu, \tau)$ is actually a mask function with the sample set $\Omega$ corresponding to the sampling rectangular area centered around the AF domain origin
\begin{align}
\mathbf{\Phi}(\nu, \tau)= \left\{ 
\begin{array}{lr}
1, & (\nu, \tau) \in \Omega \\
0, & \text{otherwise}
\end{array}
\right . \label{eq6}
\end{align}
which has totally $M=D_{\nu} \times D_{\tau}$ samples, where $D_{\nu}$ and $D_{\tau}$ denote the two dimensions of the sampling rectangle.

According to the above analysis and the sparse sampling in (\ref{eq5}), the expression in (\ref{eq4}) is reformulated as a sparse representation
\begin{align}
\boldsymbol{a}^{\prime}=\mathbf{\Psi}^{\prime}\boldsymbol{\omega},\label{eq7}
\end{align}
where $\boldsymbol{a}^{\prime} \in \mathbb{R}^{M \times 1}$ is the CS observation vector, and $\mathbf{\Psi}^{\prime} \in \mathbb{R}^{M \times N^{2}}$ is the CS measurement matrix consisting of certain Fourier transform vectors which are relevant to $\mathbf{\Phi}(\nu, \tau)$. Note that when the size of $\Omega$ is much less than that of  $\boldsymbol{\omega}$, i.e., $M \ll N^{2}$, the expression in (\ref{eq7}) can be viewed as a classical CS sparse reconstruction problem by solving the following LASSO problem \cite{lasso1996}
\begin{align}
\widehat{\boldsymbol{\omega}}={\arg\underset{\boldsymbol{\omega}}\min} \frac{1}{2}||\boldsymbol{a}^{\prime}-\mathbf{\Psi}^{\prime}\boldsymbol{\omega}||_{2}^{2}+\lambda||\boldsymbol{\omega}||_{1}, \label{eq8}
\end{align}
where $\lambda>0$ is a regularization parameter. The objective of the proposed U-ISTA is to learn the CT-free $\boldsymbol{\omega}$ in  (\ref{eq8}), which is expected to be close to ideal TF representation.

\subsection{U-ISTA Network Architecture}
An efficient method for solving the problem in (\ref{eq8}) is ISTA, of which each iteration includes matrix-vector multiplication involving $\mathbf{\Psi}^{\prime}$ and $\mathbf{\Psi}^{\prime T}$, followed by a shrinkage-thresholding step \cite{daubechies2004iterative}
\begin{align}
\boldsymbol{\omega}^{(k+1)}&=h_{\theta} \left(\boldsymbol{\omega}^{(k)}-\frac{1}{L}\mathbf{\Psi}^{\prime T} \left(\mathbf{\Psi}^{\prime}\boldsymbol{\omega}^{(k)}-\boldsymbol{a}^{\prime} \right)\right), \label{eq9}
\end{align}
where the soft shrinkage function is denoted as $h_{\theta}(\boldsymbol{\omega})=\sgn({\boldsymbol{\omega}})(|{\boldsymbol{\omega}}|-\theta)$,  where $\theta=\lambda / L $ is a fixed threshold, and $L$ is the largest eigenvalue of $\mathbf{\Psi}^{\prime T}\mathbf{\Psi}^{\prime}$. The above shows that all the thresholds in standard ISTA merely depend on regularization parameter $\lambda$ and the largest eigenvalue $L$. Extended researches related to ISTA pay increasing attention to proper selection of thresholds \cite{7481717,fosson2018biconvex}. In addition, learning all parameters is considered in Learned ISTA (LISTA) \cite{gregor2010learning}, thus theoretical fast convergence rate as well as lower error rates are gained \cite{chen2018theoretical,liu2018alista}. 

However, most of existing supervised algorithms pay less attention to  the inter-dependencies between signal coefficients, e.g., the TF continuity of analyzed signal. Inspired by the fact that exploring TF structured sparsity will lead to better signal reconstruction, we propose a new U-Net aided ISTA network architecture, as illustrated  in Fig. \ref{fig1}. At first, we unfold the ISTA as a resemble recurrent neural network (RNN) like the LISTA \cite{gregor2010learning}, for which it is more convenient to introduce structured sparse prior.
Specifically, we make a choice of fixing $\mathbf{\Psi}^{\prime}$ in every layer other than learning it through data since Fourier transform matrix has an explicit physical significance. Furthermore, with a proper selection of threshold, i.e., making small thresholding values in nonzero entries while setting large values in elsewhere, higher convergence rate and lower possible error will be achieved \cite{chen2018theoretical}. 

Resorting to the power of advanced deep learning technology, the structure-aware thresholds could be learned  benefiting from sufficient training data. For the sake of exploiting the continuity structure behind sparsity, we need to expand effective receptive field so as to attain long-range spacial dependencies among coefficients of the signal's TFD. To achieve this, the typical U-Net classifier, which combines shallow with semantic features to assign a label to each pixel, is employed to explore TF correlation structure of the signal, e.g., structure-aware thresholding values are prone to be larger in ATs while smaller values are learned in CTs. Additionally, we employ only three layers in encoder and decoder path to balance between  complexity and high-resolution, in the meanwhile, channels are considerably reduced.

In the framework of U-ISTA, the solution to the problem in (\ref{eq8}) is written as
\begin{equation}
\label{eq9}
\left\{
\begin{aligned}
 \boldsymbol{\omega}^{(k+1)} &= h_{\theta^{(k)}}\big(\mathbf{P}^{(k)}\boldsymbol{\omega}^{(k)}+\mathbf{Q}^{(k)}\boldsymbol{a}^{\prime}\big)\\ 
 \mathbf{P}^{(k+1)} &=\mathbf{I}-t^{(k+1)}\mathbf{\Psi}^{\prime T}\mathbf{\Psi}^{\prime}\\
 \mathbf{Q}^{(k+1)} &=t^{(k+1)}\mathbf{\Psi}^{\prime T}
\end{aligned}
\right.
\end{equation}
where $k=0,1,2,...,K-1$, resulting in a $K$-layer network, $t^{(k)}$ is a scalar which is the key to the convergence rate of network \cite{liu2018alista}, and $\mathbf{I}$ denotes identity matrix. Hence, the structure-aware threshold is defined as
\begin{align}
\theta^{(k)}=U\left\{\boldsymbol{\omega}^{(k)}\right\}, \quad k=0,1,2,...,K-1\label{eq10}
\end{align}
where $U\left\{\cdot\right\}$ represents the mapping from the $k^{th}$ iteration signal TF reconstruction  to the structure-aware thresholds at the $k^{th}$ U-Net block. Generally, in order to achieve a near-ideal TF representation, the proposed U-ISTA adaptively controls thresholds  according to the analyzed signal structure by learning local dependencies among signal amplitudes, which further makes a compensation for simple sparsity.

\section{Numerical Experiments}

We train the proposed U-ISTA network on synthetic non-stationary signal data for 2500 epochs using NVIDIA GeForce GTX 1080 GPUs with 0.001 learning rate. To be specific, the number of iterations in (\ref{eq10})  is set to $K=5$, and the training data contain various signal mixtures consisting of randomly matched linear frequency-modulated (LFM) and nonlinear sinusoidal frequency-modulated (SFM) components, with SNR levels ranging from 5 dB to 25 dB. Furthermore, the smooth $\ell_{1}$ loss function is employed to obtain a robust network. For all reconstruction results, amplitudes are coded logarithmically with a dynamic range of 20 dB. The U-ISTA takes $29\times 29$ sampling rectangular area while the $\ell_{1}$-app algorithm merely samples $13 \times 13$ points \footnote{The performance of $\ell_{1}$-app degrades as the size of sampling area increases, thus the number of samples in $\ell_{1}$-app is smaller than that of U-ISTA to reduce the probability of  sampling CTs in AF domain.}. The proposed U-ISTA  is compared against some commonly used state-of-the-art algorithms, including BD \cite{950779}, WVD \cite{2016ivBoualem}, $\ell_{1}$-app \cite{flandrin2010time}, and RSP \cite{382394} \footnote{We would like to thank the authors in these works for sharing source codes, which are available online.}. In our simulation, the averaged normalized mean square error (NMSE) is used as performance assessment criterion
\begin{align}
\mathrm{NMSE}=10 \log _{10} \left\{\frac{\|\boldsymbol{\omega}-\hat{\boldsymbol{\omega}}\|_{2}^{2}}{\|\boldsymbol{\omega}\|_{2}^{2}}\right\}.
\end{align} 

In the following, we consider five different cases with both linear LFM component as well as nonlinear SFM component, and each component has $N=128$ time samples. The visualization of the five non-stationary mixtures in 2-D TF domain is provided in the first columns of Fig. \ref{fig2} and Fig. \ref{fig3}, which examine various situations including complex conditions, e.g., closely-located and spectrally-overlapped components. To verify the effectiveness of different algorithms on varying signal amplitudes, an amplitude modulated (AM) function is imposed on each component. As shown in Fig. \ref{fig2} and Fig. \ref{fig3}, the reconstructed TFDs by different recovery algorithms are presented in favorable noise environment (SNR = 45 dB) and adverse noise environment (SNR = 5 dB), respectively.

\vspace{2mm}
\subsection*{\textbf{Case 1}: signal composed of two far-located LFMs}
Considering a mixture signal which contains two far-located LFM components:
\begin{align}
\begin{cases}
x_{1}(t)=a(t)\cos \left\{ 2\pi \left( 0.0002(t^{2}-t_{0}^{2})+0.441(t-t_{0}) \right) \right\}\\
x_{2}(t)=a(t)\cos \left\{ 2\pi \left( 0.0004(t^{2}-t_{0}^{2})+133(t-t_{0}) \right) \right\}
\end{cases}\notag
\end{align}
where $a(t) = \exp\left\{{-(0.0016t - 1)^{2} \pi}\right\}$ is the AM function, and $t_{0}$ equals to 64 in all five cases.

When SNR = 45 dB, it is seen from the first row of Fig. \ref{fig2}  that the CTs are significantly suppressed by $\ell_{1}$-app, RSP and U-ISTA, while BD and WVD suffer from varying degrees of CT interference. It is worth mentioning that the proposed U-ISTA  achieves near-ideal TF concentration which is close to ideal TFD. 
The reconstruction results when SNR = 5 dB are given in the first row of Fig. \ref{fig3}, note that although the performance of the U-ISTA degrades in contrast to the high SNR case, the proposed U-ISTA still obtains the highest TF resolution  compared with other recovery algorithms, which are prone to be seriously disturbed by strong noise.

\subsection*{\textbf{Case 2}: signal composed of two closely-located LFMs}
Considering a mixture signal which contains two closely-located LFM components:
\begin{align}
\begin{cases}
x_{1}(t)\!=\!a(t)\cos \!\left\{ 2\pi \left( -0.0003(t^{2}-t_{0}^{2})+0.3164(t-t_{0}) \right) \right\} \\
x_{2}(t)\!=\!a(t)\cos \!\left\{ 2\pi \left( 0.0001(t^{2}-t_{0}^{2})+211(t-t_{0}) \right) \right\} 
\end{cases}\notag
\end{align}

When signal's components are closely located to each other, the CTs tend to become serious, thus posing a challenge to the existing literature. The reconstruction results in high and low SNR levels are shown in the second rows of Fig. \ref{fig2} and Fig. \ref{fig3}, respectively. It is seen that the results of U-ISTA give high TF resolution without CTs even though partly deformed when SNR = 5 dB. On the contrary, severe distortion appears in $\ell_{1}$-app while other algorithms suffer from considerable CTs and noise influence, which make them hard to analyze and interpret the mixture signal.

\subsection*{\textbf{Case 3}: signal composed of two overlapped LFMs}
Considering a mixture signal which contains two overlapped LFM components:
\begin{align}
\begin{cases}
x_{1}(t)=a(t)\cos \left\{ 2\pi \left( -0.0006(t^{2}-t_{0}^{2})+0.29(t-t_{0}) \right) \right\} \\
x_{2}(t)=a(t)\cos \left\{ 2\pi \left( 0.0011(t^{2}-t_{0}^{2})+0.109(t-t_{0}) \right) \right\} 
\end{cases}\notag
\end{align}  

\begin{figure*}[!t]
\centering  
\includegraphics[scale=0.63]{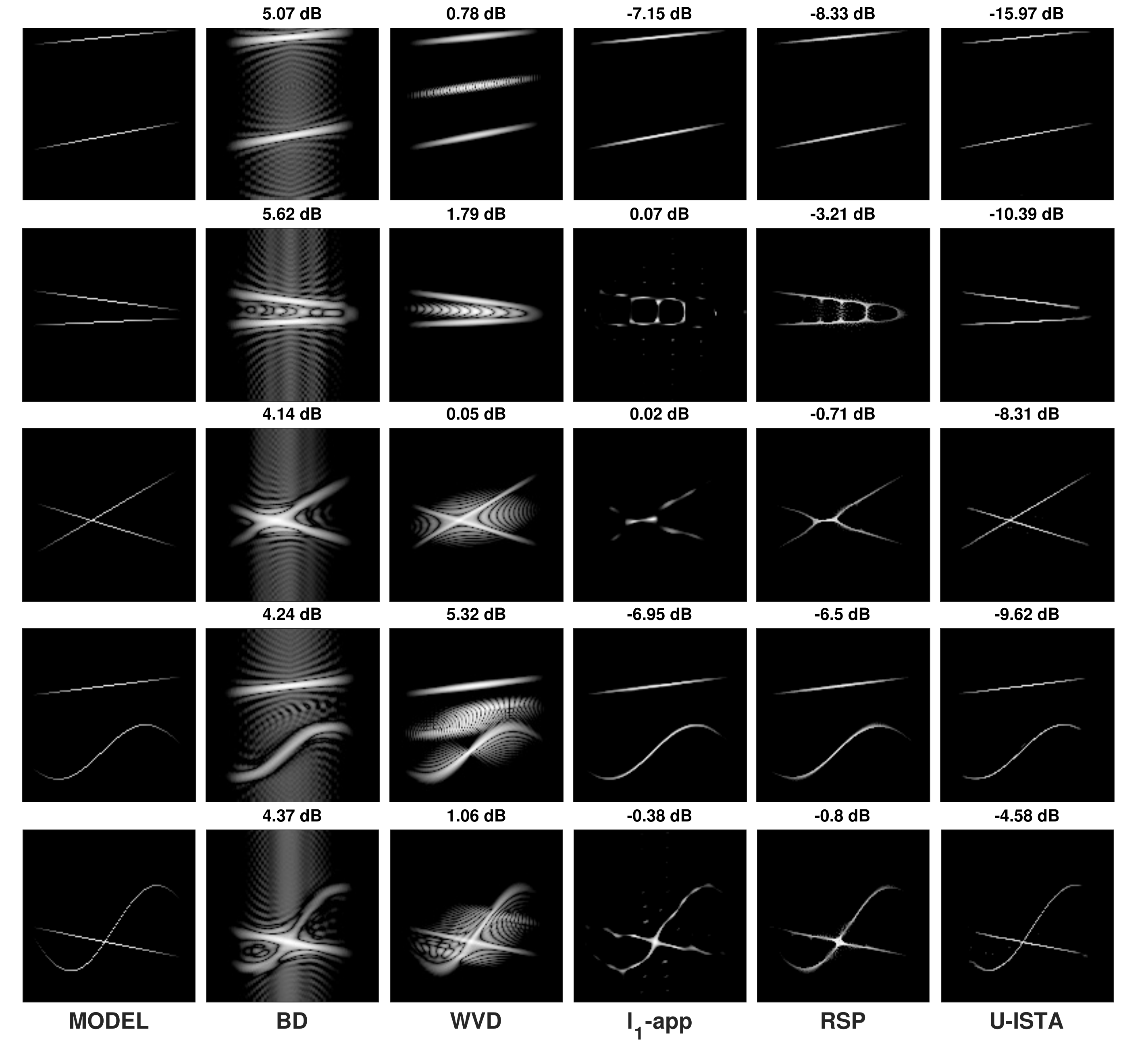}
\caption{TF reconstructions of the five cases at \textbf{SNR = 45 dB} and NMSE is annotated above each image (From top to bottom: \textbf{\textit{Case 1}} to \textbf{\textit{Case 5}}. From left to right: Ideal TFD, BD \cite{950779}, WVD \cite{2016ivBoualem}, $\ell_{1}$-app \cite{flandrin2010time}, RSP \cite{382394}, and the proposed U-ISTA).}
\label{fig2}
\end{figure*}

\begin{figure*}[!t]
\centering
\includegraphics[scale=0.63]{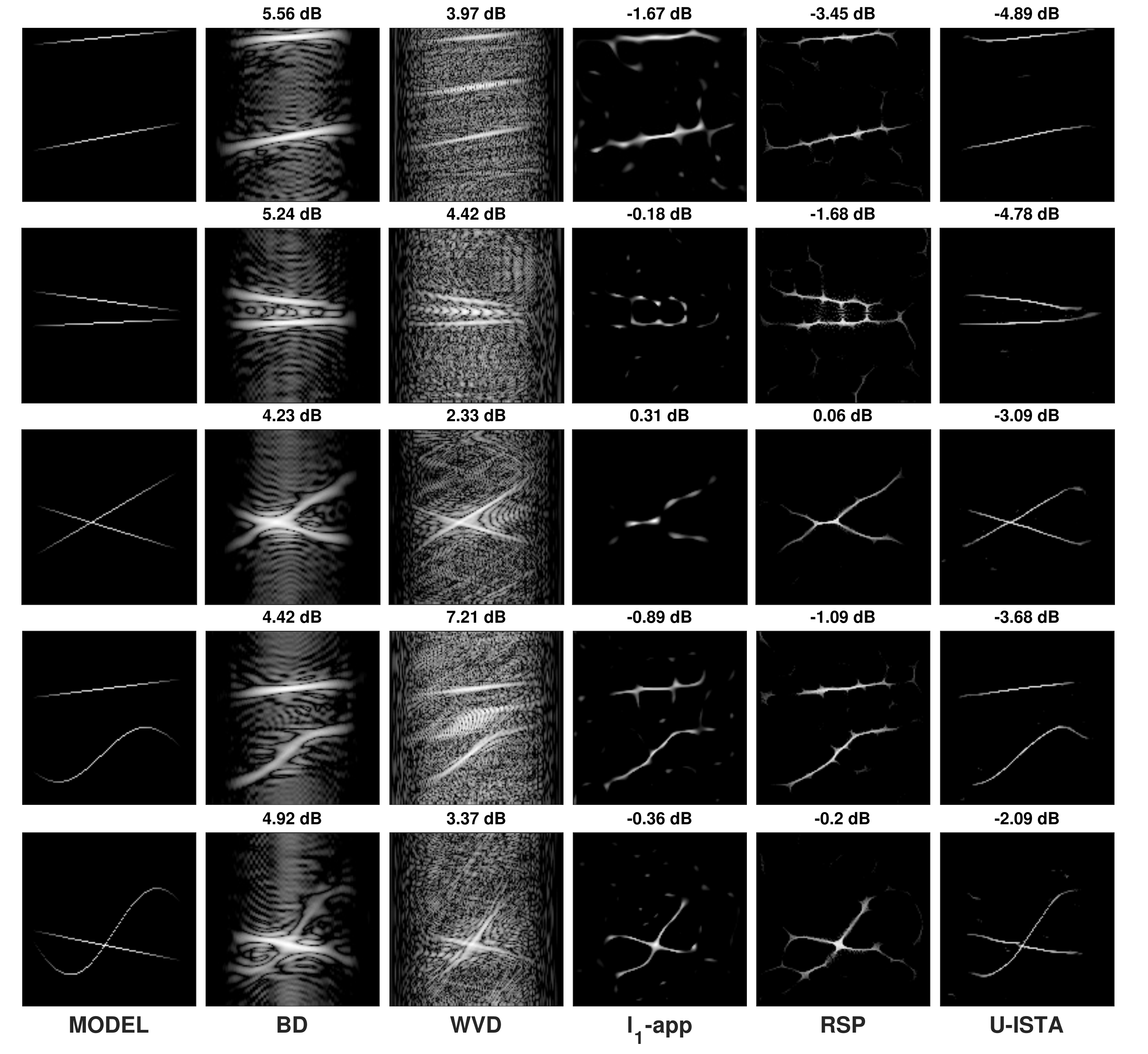}
\caption{TF reconstructions of the five cases at \textbf{SNR = 5 dB} and NMSE is annotated above each image (From top to bottom: \textbf{\textit{Case 1}} to \textbf{\textit{Case 5}}. From left to right: Ideal TFD, BD \cite{950779}, WVD \cite{2016ivBoualem}, $\ell_{1}$-app \cite{flandrin2010time}, RSP \cite{382394}, and the proposed U-ISTA).}
\label{fig3}
\end{figure*}

Recovery results are shown in the third rows of Fig. \ref{fig2} and Fig. \ref{fig3}. It is observed that the $\ell_{1}$-app fails to represent the signal since ATs and CTs are mixed up around the origin of AF domain. When SNR = 5 dB, we note that the CTs are greatly reduced for the RSP algorithm, however, the deformation of its signal TFD occurs at the intersection. Benefitting from a large number of training data,  the U-ISTA has more stable performance and succeeds to recover the signal TFD with high TF resolution. It should be emphasized that the proposed U-ISTA can accurately restore the information at the intersection, which is a difficult problem to be solved.

\begin{figure*}[!t]
\begin{center}
\subfigure[\textbf{\textit{Case 1}}: two far-located LFM components.]{\hspace{-5mm}
\includegraphics[width=9.3cm]{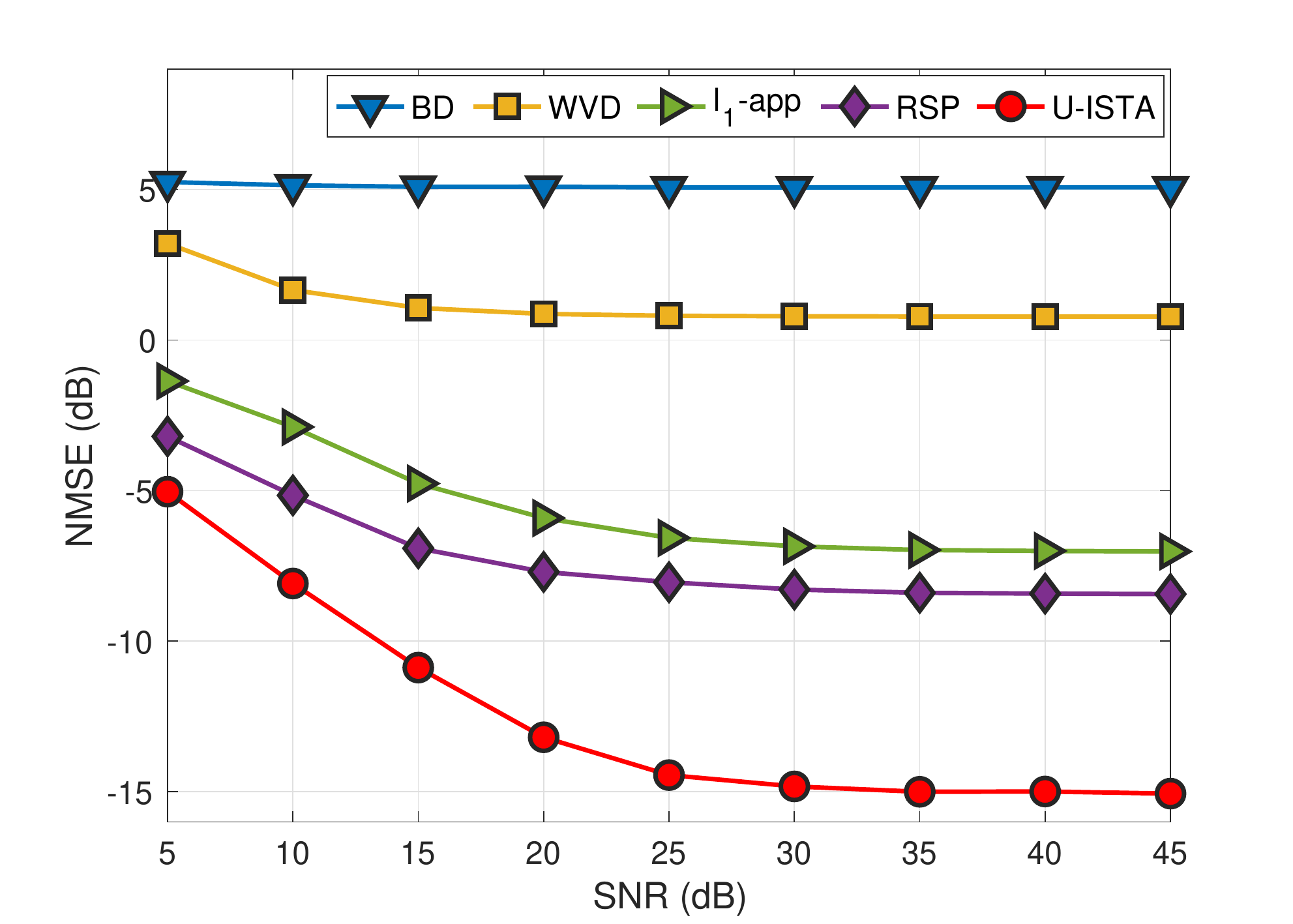}
\label{fig4c}
}
\subfigure[\textbf{\textit{Case 2}}: two closely-located LFM components.]{\hspace{-5mm}
\includegraphics[width=9.3cm]{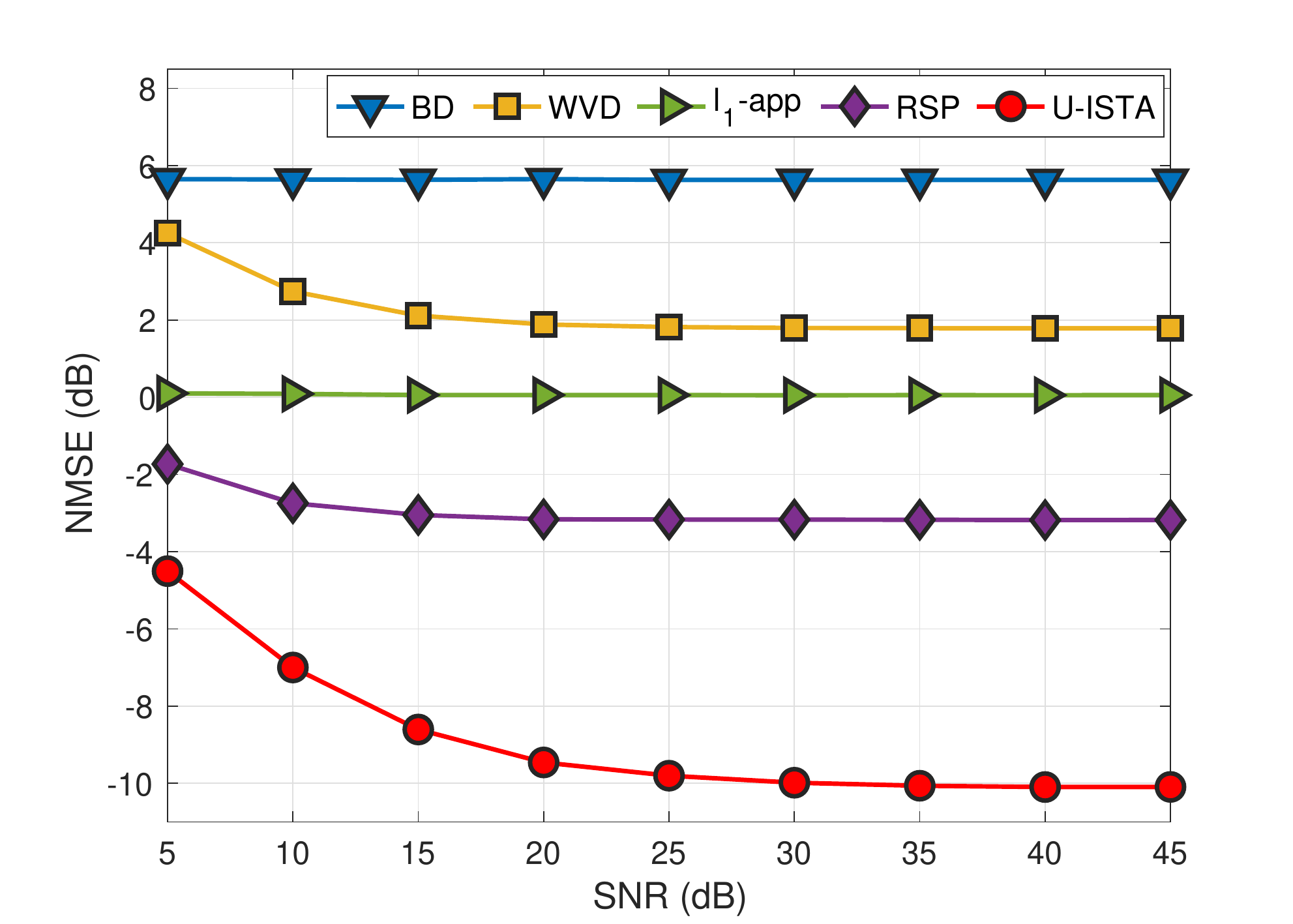}
\label{fig4c}
}

\subfigure[\textbf{\textit{Case 3}}: two overlapped LFM components.]{\hspace{-5mm}
\includegraphics[width=9.3cm]{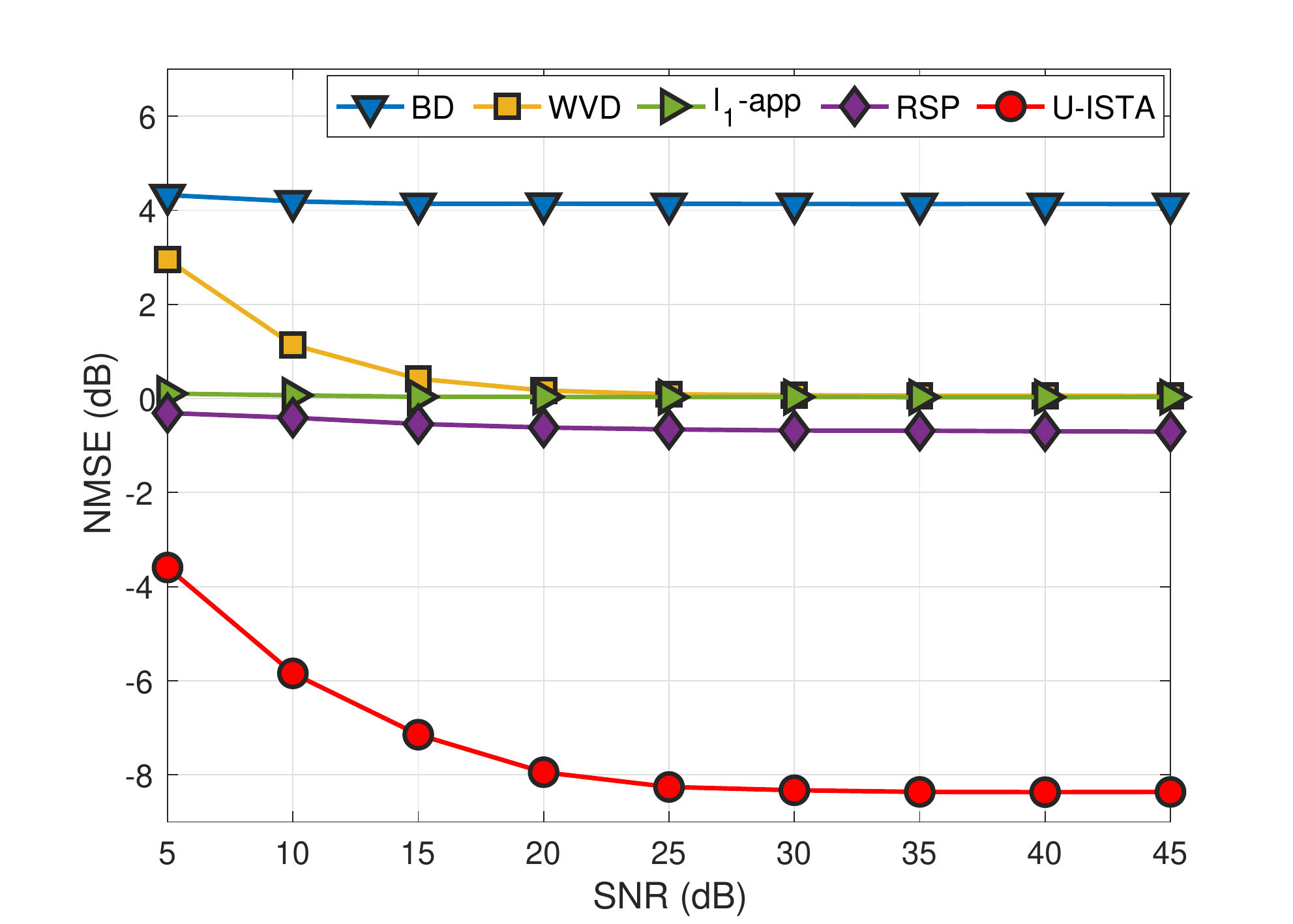}
\label{fig4c}
}
\subfigure[\textbf{\textit{Case 4}}: non-overlapped LFM and SFM components.]{\hspace{-5mm}
\includegraphics[width=9.3cm]{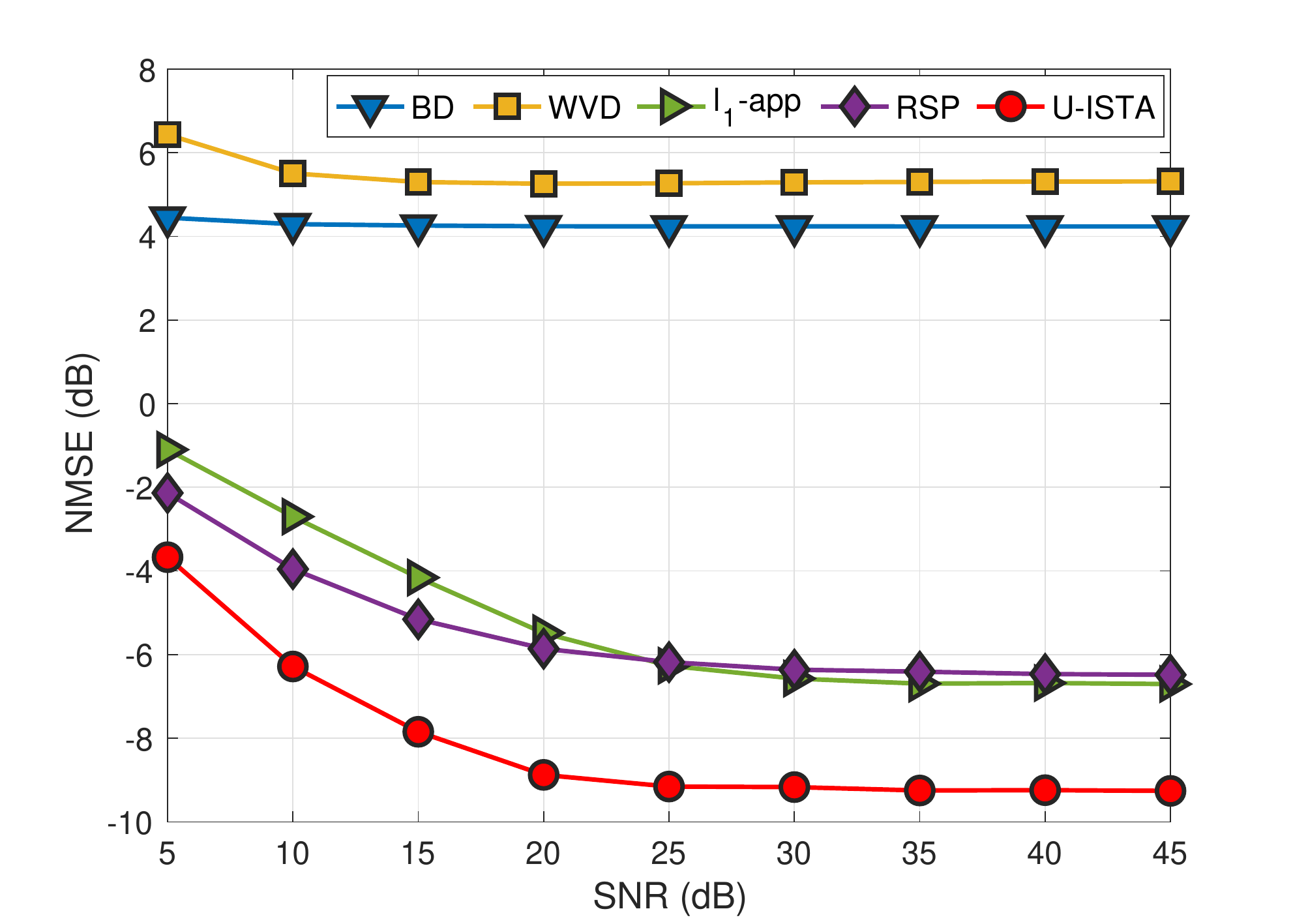}
\label{fig4e}
}

\subfigure[\textbf{\textit{Case 5}}: overlapped LFM and SFM components.]{\hspace{-5mm}
\includegraphics[width=9.3cm]{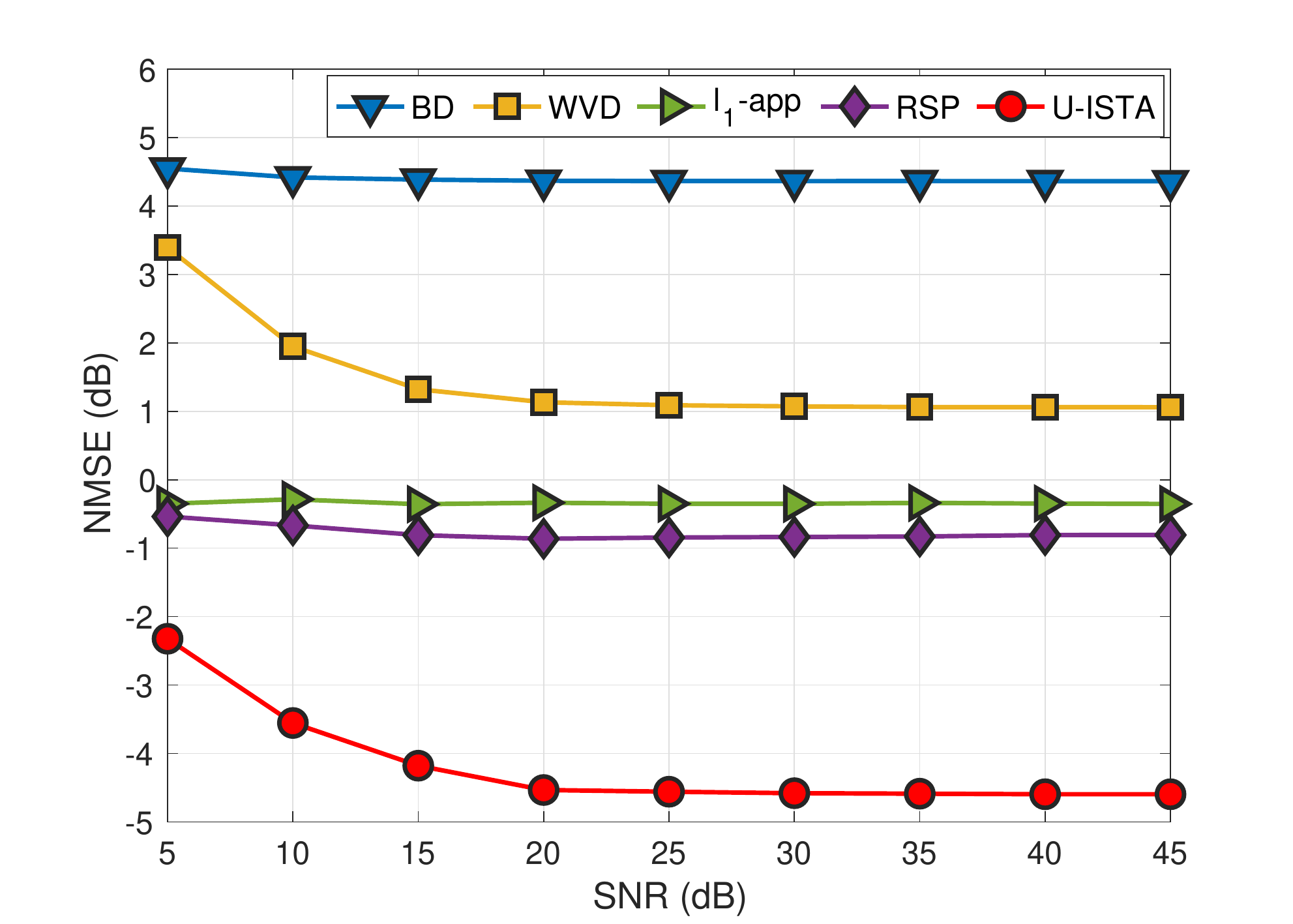}
\label{fig4c}
}
\subfigure[ Averaged NMSEs from \textbf{\textit{Case 1}} to \textbf{\textit{Case 5}}.]{\hspace{-5mm}
\includegraphics[width=9.3cm]{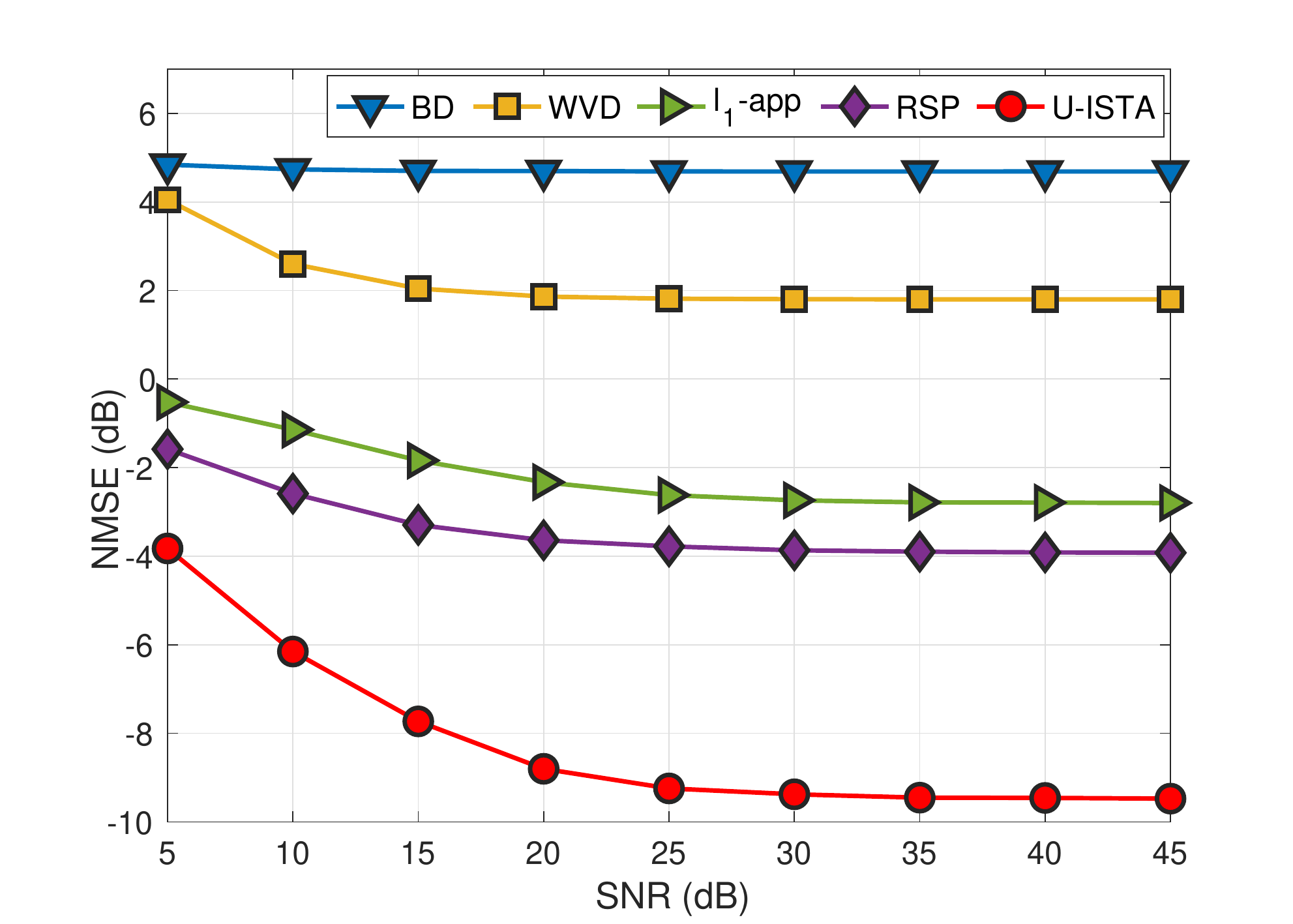}
\label{fig4c}
}
\caption{NMSE performance versus various SNR levels using different TF reconstruction algorithms: BD \cite{950779}, WVD \cite{2016ivBoualem}, $\ell_{1}$-app \cite{flandrin2010time}, RSP \cite{382394}, and the proposed U-ISTA.}
\label{fig4}
\end{center}
\end{figure*}

\subsection*{\textbf{Case 4}: signal composed of an LFM and an SFM}
Considering a mixture signal which contains an LFM component and an SFM component:
\begin{align}
\begin{cases}
x_{1}(t)\!=\!a(t)\!\cos \!\left\{ 2\pi \left( 0.0002(t^{2}\!-\!t_{0}^{2})\!+\!0.297(t-t_{0}) \right) \right\} \\
x_{2}(t)\!=\!a(t)\!\cos \!\big\{ 0.27\pi (t\!-\!t_{0}) \!\\
\!\quad\quad\quad\quad+9.98\sin\big(0.0156\pi (t-t_{0})-\phi_{0}\big)-9.98\sin\phi_{0}\big\}
\end{cases}\notag
\end{align}
where $\phi_0$  = -1.403.

In this case, we can obtain the similar observation and conclusion to those in \textbf{\textit{Case 1}}. The only difference is that the nonlinearity of the SFM component will cause inner CTs, which can be perceived in Fig. \ref{fig2}. As seen from the fourth rows of Fig. \ref{fig2} and Fig. \ref{fig3}, the $\ell_{1}$-app, RSP and U-ISTA can still gain free CTs, while BD and WVD undergo both inner and outer CTs. Furthermore, the performance of $\ell_{1}$-app and RSP algorithms  deteriorates as the SNR level decreases. In contrast, the U-ISTA preserves a superb TF concentration irrespective of signal types, which verifies the effectiveness and robustness of the proposed U-ISTA.  

\subsection*{\textbf{Case 5}: signal composed of overlapped LFM and SFM}
Considering a mixture signal which contains overlapped LFM component and  SFM component:
\begin{align}
\begin{cases}
x_{1}(t)\!=\!a(t)\!\cos \!\left\{ 2\pi \left( -0.0004(t^{2}\!-\!t_{0}^{2})+0.117(t-t_{0}) \right) \right\}\\
x_{2}(t)\!=\!a(t)\!\cos \big\{ 0.42\pi (t\!-\!t_{0}) \!\\
\quad\quad\quad~+\!15.9\sin\big(0.0156\pi (t\!-\!t_{0})-\phi_{0}\big)-0.0156\sin\phi_0 \big\}
\end{cases}\notag
\end{align}
where $\phi_0$  = -1.832.

This case is similar to \textbf{\textit{Case 3}} except that the performance of $\ell_{1}$-app and RSP at the intersection is slightly improved, as shown in the last rows of Fig. \ref{fig2} and Fig. \ref{fig3}. Note that the $\ell_{1}$-app  recovers the signal TFD with some remaining CTs and artifacts, which may  be derived from the unapt selection of sampling area.  Besides, the proposed U-ISTA  cannot completely remove the CTs, and the reason behind this  may be the oversized sampling area, i.e., $D_{\nu} \times D_{\tau}$ is  too large so that  excessive CTs are contained. It implies that selecting a more proper sampling area is worth investigating in future.
When SNR = 5 dB, the last row of Fig. \ref{fig3} illustrates that only U-ISTA preserves a high resolution TFD with CTs greatly eliminated.

The quantitative evaluation results on different recovery algorithms for  the above five cases in various SNR levels are presented in Fig. \ref{fig4} (a) to (e), where reconstruction accuracy at each SNR level is assessed via NMSE over 100 Monte Carlo runs. In addition, the averaged NMSEs over the five cases are  shown in Fig. \ref{fig4} (f).  We observe that the NMSE results are consistent with those in Fig. \ref{fig2} and Fig. \ref{fig3}. By learning structured sparsity of training data with varying SNR levels, the U-ISTA  has achieved a large performance gain compared to other algorithms in the entire SNR range from 5 dB to 45 dB.
When the SNR drops below 5 dB, the strong noise will burden the learning blocks, thus resulting in performance deterioration. It is worth stressing that although our training data merely include synthetic  signals with SNR levels ranging from 5 dB to 25 dB, the proposed U-ISTA  still gains distinguished results outside this range, e.g., SNRs varying from 30 dB to 45 dB in Fig. \ref{fig4}, which verifies its good generalization ability. 
In addition to NMSE, quantitative evaluation in terms of $\ell_{1}$-distance to model and R\'{e}nyi entropy can also be considered to compare different  algorithms \cite{flandrin2010time}.

\section{Conclusion}
In this paper, we attempt to reconstruct near-ideal TFD by proposing the data-driven U-ISTA model, which combines the unfolded ISTA with the U-Net block to learn structure-aware thresholds so that the TF correlation structure of signals is explored. The effectiveness and robustness of the proposed U-ISTA are examined via numerical experiments in terms of NMSE, demonstrating much more significant performance improvement on synthetic non-stationary data compared to state-of-the-art algorithms, including classical TFDs and CS based ones. Particularly, the  proposed U-ISTA successfully reconstructs more signal details, e.g., the intersection region of spectrally-overlapped components, while maintaining superb TF concentration and continuous characteristics of signals even for closely-located signal components. Although exposing the potential benefit by sampling in AF domain, it is believed that proper selection of AF samples would result in better reconstruction results, thus a learning strategy for optimal sampling is worth investigating for future research. In addition, we also believe that our U-ISTA model can be boosted by feeding a wide variety of synthetic data as well as real-life measured data.


\ifCLASSOPTIONcaptionsoff
  \newpage
\fi
\bibliographystyle{IEEEtran}
\bibliography{jl_v1}
\end{document}